\pdfoutput=1
\documentclass[aps,pra,twocolumn,showpacs,superscriptaddress,10pt,letterpaper,floatfix,longbibliography]{revtex4-1}

\usepackage[utf8]{inputenc} 
\usepackage[T1]{fontenc}	
\usepackage{geometry}		
\usepackage{graphicx}
\usepackage[above,below]{placeins}	
\usepackage{times}
\usepackage{multirow}
\usepackage{hyperref}
\usepackage{amsmath}
\usepackage{cleveref}
\usepackage{enumitem}
\usepackage{longtable}

\usepackage{hyperref}  
\hypersetup{colorlinks=true,urlcolor=blue,citecolor=blue,linkcolor=blue}   
\usepackage{enumitem}          
\setlist{nosep}                 

\begin{document}


  \title{Rubric-based holistic review: a promising route to equitable graduate admissions in physics}

  \author{Nicholas T. Young}
  \email[Current email: ]{ntyoung@umich.edu}
  \affiliation {Department of Physics and Astronomy, Michigan State University, East Lansing, Michigan 48824}
  \affiliation {Department of Computational Mathematics, Science, and Engineering, Michigan State University, East Lansing, Michigan 48824}
 
 \author{K. Tollefson}
 \affiliation {Department of Physics and Astronomy, Michigan State University, East Lansing, Michigan 48824}
 
  \author{Remco G. T. Zegers}
  \affiliation {Department of Physics and Astronomy, Michigan State University, East Lansing, Michigan 48824}
  \affiliation{National Superconducting Cyclotron Laboratory, Michigan State University, East Lansing, Michigan, 48824}
  \affiliation{Joint Institute for Nuclear Astrophysics, Michigan State University, East Lansing, Michigan, 48824}

  \author{Marcos D. Caballero}
  \email[Corresponding Author: ]{caball14@msu.edu}
  \affiliation {Department of Physics and Astronomy, Michigan State University, East Lansing, Michigan 48824}
  \affiliation {Department of Computational Mathematics, Science, and Engineering, Michigan State University, East Lansing, Michigan 48824}
  \affiliation {Center for Computing in Science Education \& Department of Physics, University of Oslo, N-0316 Oslo, Norway}
  \affiliation {CREATE for STEM Institute, Michigan State University, East Lansing, Michigan 48824}


  \begin{abstract}
    As systematic inequities in higher education and society have been brought to the forefront, graduate programs are interested in increasing the diversity of their applicants and enrollees. Yet, structures in place to evaluate applicants may not support such aims. One potential solution to support those aims is rubric-based holistic review. Starting in 2018, our physics department implemented a rubric-based holistic review process for all applicants to our graduate program. The rubric assessed applicants on 18 metrics covering their grades, test scores, research experiences, noncognitive competencies, and fit with the program. We then compared faculty’s ratings of applicants by admission status, sex, and undergraduate program over a three-year period. We find that the rubric scores show statistically significant differences between admitted and non-admitted students as hoped and that statistically significant differences based on sex or undergraduate program aligned with known disparities in GRE scores and service work expectations. Our results then suggest rubric-based holistic review as a possible route to making graduate admissions in physics more equitable.
  \end{abstract}

  \maketitle

\section{Introduction}\label{sec:introduction}
Female and Black, Latinx, and Indigenous scholars have been and are underrepresented at all levels of physics. The percentage of physics degrees awarded to women has stagnated at around 20\% \cite{patrick_j_mulvey_trends_2021} while the percentage of physics degrees awarded to Black, Latinx, and Indigenous students has remained less than 10\% despite these students making up a larger portion of the college population than in the past \cite{hodapp_making_2018}. While there are numerous possibilities to address the systematic inequities these scholars face at all levels of academia that limit their participation \cite{the_aip_national_task_force_to_elevate_african_american_representation_in_undergraduate_physics__astronomy_time_2020,rybarczyk_analysis_2016,sensoy_we_2017,eisen_model_2017,bhalla_strategies_2019,cardel_turning_2020,posselt_equity_2020}, this paper will focus on graduate admissions in physics. Specifically, if we treat graduate admissions as a four stage process similar to how O'Meara et al. treats faculty hiring as a four-stage process \cite{omeara_nudging_2020}, this paper focuses on the evaluating applicants and making admissions offers stages of the process.

While physics departments may be interested in increasing their diversity, the dominant processes of evaluating applicants for graduate school do not support such aims. Prior work has found that diversity considerations are often secondary when evaluating applicants and are discussed after many diverse candidates have already been cut from the applicant pool \cite{posselt_toward_2014,posselt_metrics_2019}. Therefore, increasing diversity and equity during the admissions process requires rethinking the process physics departments use to evaluate applicants.

One promising approach to rethinking the admissions process is holistic review, where a broad range of candidate qualities are considered \cite{kent_holistic_2016}. In physics, the use of rubric-based review to facilitate such holistic reviews has been gaining traction through the Inclusive Graduate Education Network \cite{noauthor_inclusive_nodate}. Under this approach, applicants are rated on both traditional metrics such as GPA and test scores as well as noncognitive skills such as showing initiative and displaying perseverance according to a predefined rubric. Such an approach is claimed to ensure that each applicant is treated fairly and biases by reviewers are checked \cite{miller_equitable_2020}, and hence, it could make graduate admissions more equitable.

To our knowledge, however, few studies have examined how these rubrics work in practice and whether they fulfill such aims. Therefore, the goal of this paper is to examine empirically those claims in the context of our department's graduate program. Specifically, our paper addresses three questions related to rubric-based review in our department:

\begin{enumerate}
    \item How do faculty assign rubric scores to applicants and how do those differ between admitted and rejected applicants?
    \item How do the scores assigned by faculty differ by applicant’s sex?
    \item How do the scores assigned by faculty differ by the type of institution the applicant attended?
\end{enumerate}

As Scherr et al. concluded in their study of graduate admissions practices in physics, many departments are unaware of what other departments do and hence, they might be willing to change their practices if they become aware of successful practices in use elsewhere \cite{scherr_fixed_2017}. Therefore, a secondary goal of this paper is to describe alternative admissions practices in physics and how departments may apply these alternative practices to their own admissions processes.

The rest of the paper is organized as follows. In Sec. \ref{sec:background}, we provide an overview of holistic review, rubric-based review, and evidence from other fields about their potential for success. In Sec. \ref{sec:methods}, we describe how our department transitioned to rubric-based review, how we collected data relevant to evaluating our admissions process, and how we analyzed such data. In Sec. \ref{sec:Results}, we share results that suggest our rubric does support more equitable admissions practices and in Sec. \ref{sec:discussion}, we contextualize our results and examine how our choices as researchers may affect the results. In Sec. \ref{sec:Limitations} and Sec. \ref{sec:future_work} we examine the limitations of this study and suggest directions for future work. Finally, in Sec. \ref{recommendations}, we provide recommendations for departments interested in adopting rubric-based review.

\section{Background}\label{sec:background}
\subsection{A typical admissions process in physics}
When applying to a physics graduate program in the United States, an applicant will submit their undergraduate transcripts, general and physics GRE scores, multiple statements addressing their background, prior preparation, and research interests, and letters of recommendation. A group of physics faculty, the admissions committee, then reviews the applications and offers admission to some of the applicants.

Historically, there have been two main approaches for admitting students: emphasizing research or emphasizing grades \cite{doyle_search_2015}. More recent work however has tended to find that programs, including the one studied in this paper, emphasize grades and test scores over research, both in terms of what faculty say they do \cite{potvin_investigating_2017,chari_understanding_2019} and what faculty actually do \cite{posselt_inside_2016, young_using_2020}.

Numerous potential equity issues emerge when admissions is focused around test scores and grades. First, there is evidence that GRE scores vary based on gender and race \cite{miller_test_2014,miller_typical_2019} and the type of undergraduate university the test-taker attends \cite{mikkelsen_investigating_2021}. When combined with the practice of using cutoff scores, which Potvin et al. estimate at least 1 in 3 departments do despite the creators of the GRE and physics GRE recommending against it \cite{potvin_investigating_2017}, applicants from underrepresented groups in physics may be more likely to not make the first cut.

Second, the tests themselves can be a financial burden for students \cite{cochran_identifying_2018}. The cost to take the General GRE is currently \$205 in most parts of the world (and up to \$255 in some regions) \cite{noauthor_gre_nodate} and the cost to the take the physics GRE is \$150 \cite{noauthor_gre_nodate-1}. In addition, if the applicant applies to more than 4 programs, they must pay \$27 per school to send their scores. As Owens et al. notes, some students also need to travel to a testing center, which may incur travel or lodging costs \cite{owens_physics_2020}.

Third, grades vary by applicants' demographics and the type of university they attended. Whitcomb and Singh found that wealthier, continuing-generation, white students earned higher grades and that even the most privileged racially-underrepresented students in physics earned lower grades than the least privileged white students \cite{whitcomb_not_2020}. Additionally, grades are not standardized measures across universities, with students at private universities tending to be awarded slightly higher grades than their peers at public universities \cite{rojstaczer_where_2012}.

Further, evidence has not necessarily supported these metrics as useful predictors of who will earn their PhD. For example, Miller et al. found that while grade point averages were useful to some degree for predicting completion, the physics GRE had limited use \cite{miller_typical_2019}. More recent evidence suggests that the physics GRE and undergraduate grade point average only have a relation to PhD completion because they are related to graduate grade point average, which is then related to PhD completion \cite{verostek_admissions_2021}.

Given known issues with test scores and GPA, why do programs continue to emphasize them over the qualitative parts of the application. Perhaps the simplest answer is that comparing numbers is quick and convenient \cite{posselt_inside_2016}. A more nuanced answer might be that qualitative parts of an application can contain substantial variability in what is addressed and these parts of an application can have their own inequities (see. Woo et al. for an overview \cite{woo_bias_2020}).

One possible conclusion is then that all application materials have inequities, after all they are produced in an inequitable society, so what is the point of changing anything. We instead adopt a pragmatic view that some parts of the admissions process are more inequitable than others and, therefore, our goal is to develop methods to minimize or eliminate inequities to the best of our ability in an inequitable society.

\subsection{Holistic Review}
One possible approach to addressing inequities in the admissions process is holistic review, which Kent and McCarthy define as "the consideration of a broad range of candidate qualities including `noncognitive' or personal attributes" \cite{kent_holistic_2016}. Here, we will use holistic review to refer to the general process regardless of what tools or systems are used to conduct it. When talking about our department's rubric-based process or similar processes, we will use rubric-based holistic review.

While the idea of holistic admissions is hardly new, its implementation is becoming more common due to both greater awareness that quantitative measures may not accurately predict success in graduate school \cite{petersen_multi-institutional_2018,hall_predictors_2017,sealy_gre_2019} and institutions wanting to use the most predictive measures of success in their programs \cite{kent_holistic_2016}. In addition, professional societies such as the American Astronomical Society (AAS) have called for programs to implement "evidence-based, systematic, holistic approaches" to graduate admissions \cite{rudolph_final_2020}.

Using holistic review has also been claimed to lead to beneficial outcomes for universities including increasing diversity and improving student outcomes (see \cite{kent_holistic_2016}), though most of these studies have happened outside of physics and related fields. For example, Hawkins found that using holistic review increased diversity in a Doctor of Physical Therapy program \cite{hawkins_impact_2020} and in a literature review of predominantly medicine-related fields, Francis et al. found that holistic review generally increased racial and ethnic diversity \cite{francis_holistic_2021}. For STEM fields, Wilson et al. found that using holistic review in a biomedical science program resulted in applicant assessments that were independent of gender, race, and citizenship status \cite{wilson_model_2019} and Pacheco et al. found that using a composite score that included GPA, test scores, research experience, and publications was correlated with earning a university fellowship and a shorter completion time while applicant's test scores and GPAs individually were not \cite{pacheco_beyond_2015}.

While holistic review shows promise, programs may have concerns about implementing it. For example, common concerns include limited faculty time to review applications, a lack of data correlating admissions criteria and student success, and limited resources to implement it \cite{kent_holistic_2016}. In addition, there may be concerns that because the decisions can be more subjective than using a quantitative measure like a test score, there may be variability based on who reviews the application. However, a study of holistic admissions at the undergraduate level found that only 3\% of reviews showed substantial variability in the overall score between reviewers \cite{moody_rideout_study_2017}, suggesting that in practice variability in the overall rating between reviewers is limited.

\subsubsection{Noncognitive skills}
Regardless of the specifics of a holistic review process, most approaches include some examination of the applicant's noncognitive skills, which may also be referred to as soft skills, personality traits, character traits or socio-emotional skills depending on the discipline or context \cite{kautz_fostering_2014}. While there are multiple definitions of these (see \cite{roberts_back_2009}) we adopt Roberts' definition that noncognitive skills or personality traits are "the relatively enduring patterns of thoughts, feelings, and behaviors that reflect the tendency to respond in certain ways under certain circumstances" \cite{roberts_back_2009}. Often these have been operationalized as the Big Five, which are openness to experience, conscientiousness, extroversion, agreeableness, and neuroticism \cite{almlund_personality_2011,kautz_fostering_2014}, though other categorizations exist. For example, in higher education admissions, Sedlacek proposed eight noncognitive traits, which he defines as things not measured by standardized tests: positive self-concept, realistic self-appraisal, understands and knows how to handle racism (the system), prefers long-range to short-term or immediate needs, availability of strong support person, successful leadership experience, demonstrated community service, and knowledge acquired in or about a field \cite{sedlacek_noncognitive_2010}.

In terms of their utility, noncognitive skills have been found to be predictive or correlated with academic success, though these studies have happened outside of the context of physics. At the undergraduate level, noncognitive skills in isolation and in concert with test scores have been found to be more predictive of success and graduation than test scores alone \cite{tracey_noncognitive_1984,tracey_prediction_1987, medrinos_beyond_2014}. Likewise, at the graduate and professional levels, noncognitive skills have been found to be correlated with GPA and class rank \cite{carp_relationship_2020,stolte_reliability_2003}, clinical performance \cite{victoroff_what_2013}, and overall success in programs \cite{peskun_effectiveness_2007,burmeister_correlation_2014} but were not found to be associated with doing well on a licensing exam \cite{moruzi_validity_2002}. Of the individual noncognitive skills, conscientiousness has been found to be most strongly and consistently associated with academic success \cite{oconnor_big_2007}.

In addition to their benefits related to academic success, noncognitve skills can be useful for promoting equity in admissions. For example, including noncognitive skills can increase diversity without harming validity \cite{kyllonen_noncognitive_2005,ployhart_diversityvalidity_2008} as noncognitive measures have been shown to be just as valid for majoritized and minoritized groups \cite{sedlacek_why_2004,kyllonen_noncognitive_2005,miller_using_2015}. While including noncognitive skills as part of admissions may seem like a hard ask of faculty, many faculty already acknowledge the usefulness of noncognitive skills in graduate school \cite{kyllonen_noncognitive_2005}, including in physics \cite{owens_identifying_2019}.

Yet, a pressing concern is how to measure such noncognitive skills accurately. While applicant self-reports or recommender ratings are typical approaches, such methods may result in inflated or skewed ratings \cite{kyllonen_noncognitive_2005}. A recent study suggests that even sharing descriptions of noncognitive skills and why they are useful for predicting later success can artificially inflate judgements \cite{chen_sensitivity_2020}. Thus, how best to measure such skills is still an active area of inquiry \cite{noauthor_equity_2021}.

\subsubsection{Rubric-Based Review}
One promising approach to implementing holistic review is rubric-based review. Under this approach, applicants are evaluated based on a set of pre-defined criteria. By pre-selecting criteria, what is required for admission is clear to reviewers and provides a structure to assess all applicants \cite{miller_equitable_2020,posselt_inside_2016}. This explicitness has been shown to enhance both validity and reliability \cite{salvatori_reliability_2001,zeeman_validity_2017, woo_bias_2020}.

In addition, rubrics can help make the admissions process more equitable \cite{posselt_inside_2016}. By explicitly laying out the review criteria and what is required to achieve each level of the rubric, all applicants can be judged fairly and individual reviewer's expectations can be mitigated \cite{miller_using_2015}. From research into other areas of academic hiring, we know that gender and racial biases exist in the hiring process, including in physics \cite{moss-racusin_science_2012, eaton_how_2020}. Specifically in graduate admissions, faculty, including astronomy and physics faculty, have been documented showing preferences to applicants with similar backgrounds as themselves or within the same research subfield of their discipline \cite{posselt_inside_2016}. Thus, rubrics offer a possible route to counter those biases. Indeed, a recent study in admissions for a psychiatry residency program found that using rubric-based holistic review led to more underrepresented applicants receiving an offer to interview compared to the traditional approach \cite{barcelo_reimagining_2021} while a recent study of grade-school writing found that teachers rated writing attributed to a Black author lower than when it was attributed to a white author but did not find the effect when the teachers were instructed to use a clearly defined rubric \cite{quinn_experimental_2020}.

As rubric-based approaches to admission are still relatively new, best practices are still in development. Yet, a few recommendations do exist \cite{miller_equitable_2020}. First, criteria should be selected before reviewing any applications with individual programs deciding what qualities are critical for success in their program \cite{rudolph_final_2020}. Second, rubrics should be coarse-grained in that there are fewer possible scores for each construct such as low, medium, or high instead of 1-10 to limit disagreements over scores \cite{miller_using_2015}. Third, each level of the rubric should be clearly defined so that a reviewer can easily determine which score an applicant should get on each construct. These levels should be picked so that each possible score will be received by many applicants \cite{miller_equitable_2020}. Finally, these criteria and levels should allow for diverse forms of excellence to be counted as achievements so that applicants with non-traditional markers of excellence are not excluded \cite{razack_beyond_2020}. For example, assessing an applicant's research abilities should go beyond their number of publications or number of years working in a research lab and could instead focus on what the applicant accomplished in the lab or what skills they have.

While rubric-based approaches have received little research in physics, they have been successfully incorporated into larger physics graduate program initiatives. Two of the most well-known initiatives are the Fisk-Vanderbilt program, which graduates one of the largest classes of Black PhD physicists in the nation \cite{stassun_fisk-vanderbilt_2011}, and the APS Bridge Program, which has successfully admitted and retained graduate students of color at rates higher than the national average \cite{hodapp_making_2018}. Even though rubrics in admission were one of many changes made, these programs suggest that rubric-based review has promise.

For a more in depth review about equitable admissions practices in STEM doctoral programs, we refer the reader to Roberts et al. \cite{roberts_review_2021}

\section{Methods}\label{sec:methods}

\subsection{Our Rubric and Applicant Evaluation Process}
In 2018, the Department of Physics and Astronomy at Michigan State University introduced a rubric-based approach to evaluate applications to the graduate program in physics, informed by the Council of Graduate Schools' 2016 report on Holistic Review in Graduate Admissions \cite{kent_holistic_2016}.  The main goal was to improve the identification of strong candidates for the program and to make the selection more equitable, thereby increasing the participation of students from underrepresented groups in the department. In preparation for the introduction of the rubric, Casey Miller and Julie Posselt, the Inclusive Practice Hub Director and Research Hub Director, respectively of the National Science Foundation supported Inclusive Graduate Education Network \cite{noauthor_inclusive_nodate}, led a workshop with faculty who served at that time in the Graduate Recruiting Committee. This workshop resulted in a selection of five rubric categories, which each had several sub categories. Applicants are ranked with a score of either 0, 1, or 2, corresponding to low, medium, or high, for each subcategory, based on defined criteria for each score. The subcategory scores are then averaged per category and category scores summed (with weights as given below) to calculate the overall score. The categories, with subcategories in parenthesis, are:
\begin{itemize}
    \item Academic Preparation, with a weight of 25\% (Physics coursework, math coursework, other coursework, and academic recognition and honors)
    \item Research, with a weight of 25\% (Variety and duration, quality of work, technical skills, and research disposition)
    \item Non-cognitive competencies, with a weight of 25\% (Achievement orientation, conscientiousness, initiative, and perseverance) 
    \item Fit with program, with a weight of 15\% (Fit with research programs of the department, fit to research programs of specific faculty, (prior) commitment to participation in the department/school community, and advocacy for and/or contributions to a diverse, equitable, and inclusive physics community
    \item GRE scores, with a weight of 10\% (General GRE scores, and Physics GRE scores)\footnote{This category was not used in 2021, and will not be used in 2022 due to the impacts of COVID-19 on students' ability to take these tests.}
\end{itemize}
The choice of these categories and subcategories was based on the discussions in the workshop and advice from the workshop leaders, and included considerations based on experiences during previous recruiting cycles. Another consideration for the choice of the categories is a reasonably close alignment with criteria used at MSU for awarding fellowship packages to students. Therefore, the rubric scoring can also be used for selecting nominations for university fellowships. This is important because fellowship nominations are due shortly after the application deadline (January 1).   

Applications for the graduate program are submitted to MSU's central application system. All folders with a complete or near-complete application package are reviewed. The applications are divided up into several groups, which each are reviewed by different members of the graduate recruiting committee. This committee has a rotating membership with representation from faculty in all major research directions present in the department. Committee members are instructed about the use of the rubric and provided with the criteria. As part of the review process, they also sort students by their interest in research area(s). The results from the rubric scoring are compiled by the Graduate Program Director. Students whose folders are near complete, but have a ranking for which an offer is not impossible, are contacted and asked to provide the missing information. If that additional information is provided, the rubric scoring is updated. 

Subsequently, the spreadsheet is used by committee representatives from each major research area in the department to make a list of students they would like to make an offer to for a position in that specific research area. The number of students who are made an offer to depends on openings available per research area, the number of teaching assistant slots available, and the historical acceptance rates for each research area. Typically, the process results in a list of offers that will be made and a wait list for additional offers that can be made if recruiting targets are not met in the initial round of offers. In this stage of the recruiting process, the match to available positions is revisited as committee members from specific research areas are better aware than general faculty members about the recruiting needs for that year. In spite of the instructions and criteria provided to reviewers, the scoring is still somewhat subject to differences in reviewing styles and interpretation of the criteria. This is, for example, apparent in the comparison of average summed scores per reviewer. Therefore, this second stage of the review process also allows for another comparison of applications based on the rubric by a few faculty members in each research area. Because of these reasons, the list of students whom an offer will be made to, or who are put on a wait list, quite closely follows the original rubric scoring but modifications do occur. 

The whole process is organized and overseen by the Graduate Program Director with support from the Graduate Program Secretary. The Graduate Program Director also serves as the point of contact for questions about the use and interpretation of the rubric, reviews applications of likely candidates, and leads the selection of nominations for fellowships.  

The overall response from faculty who served in the recruiting committee and used the rubric has been positive, as it provides clear guidance for the review process and reduces the impact of different reviewing styles and biases to what are the most important skills applicants to a physics graduate program should have. On average, the time spent by individual committee members on reviewing the folders has not increased. Faculty reviewers have provided feedback that it would be better if applicants are first sorted by research area so that the review is done by several faculty from the relevant research areas in the first step. Given the large number of applications and the limitations of the current software used to manage applications, this could not easily be accomplished in the past. MSU is implementing new software for managing and reviewing applications, which will make presorting of applications by research area possible, leading to a considerable increase in the efficiency of the process.    

\subsection{Participants and Data Collection}
Data for this study comes from compiled records from applicants to our physics graduate program for fall 2018, 2019, and 2020. Most admissions decisions for fall 2020 had already been made before coronavirus accommodations took effect, suggesting at most minimal effects on our data.

When applying to the university, applicants submit a general university application, transcripts, test scores, a personal statement, an academic or research statement, and letters of recommendation to a central system. As the current admissions system does not allow for records to be compiled across applicants, two researchers manually extracted relevant information for this study. The researchers independently extracted data from the first 20 applications and then compared results to ensure they were interpreting the applications the same and agreeing on any conventions for reporting the data. Afterwards, the researchers independently went through the rest of the applications. Through this process, the researchers collected the applicant's demographics, grade point average, GRE scores, degrees earned or in progress, and previous institutions attended. Any information missing from the applications or entered into the application on a non-standard scale (e.g. a GPA on a non 4.0 scale or a GRE score outside of the current scoring range) was treated as missing data for the analysis.

As rubric scores are determined by faculty and are not part of the materials applicants submit, aggregated scores were then matched with individual applicants using the applicant IDs. Through this process, we collected data on 826 applicants, 511 of which were domestic applicants.

\subsection{Analysis}
Because of different application requirements and availability of institutional data for international and domestic students, we only include domestic students in our study. In addition, we only include applications sufficiently complete that faculty were able to rate and were included in the Graduate Program Director compiled records, leaving us with 321 domestic applicants for this study.

For our analysis, we were interested in how faculty rate applicants and hence, we computed the fraction of applicants in each level (low, medium, and high) of the rubric. In some cases (<5\%), faculty used a rating that was in between levels (e.g. low-medium). Because of this, we performed all subsequent analyses by first rounding up (so low-medium would become medium) and then repeating the analysis by rounding down. 

First, we computed the fraction of applicants in each level of the rubric for all applicants, all admitted applicants, and all non-admitted applicants.

Second, we compared applicants based on demographics by comparing the fraction of applicants in each bin of the rubric. While gender would be more appropriate, the application system only asks applicants about their sex and allows them to choose male or female. Thus we were only able to compare faculty ratings of males and females. We acknowledge that females is not the correct term to use, but as being female does not automatically imply being a woman, we do not believe it is appropriate to assume that someone marking female as their sex is necessarily a woman.

In terms of race, the application system does not allow applicants to enter their race or ethnicity, so we are unable to compare applicants of different races.

Finally, we compared applicants from different undergraduate backgrounds because prior work suggests the applicant's background may influence faculty's perceptions of them. For example, faculty may prefer applicants with similar backgrounds as themselves \cite{posselt_inside_2016} and may interpret grade point averages in the context of the applicant's undergraduate program, with high GPAs from more prestigious universities carrying more "weight" than a high GPA from a lesser known school \cite{posselt_trust_2018}. In addition, graduate admissions in physics have been characterized as "risk-adverse" where faculty prefer to admit applicants who are likely to complete their program rather than take chances on someone who might not \cite{posselt_inside_2016,scherr_fixed_2017}. As students from smaller programs may be viewed as higher risk if previous students from that program struggled \cite{posselt_trust_2018}, it is possible faculty may be less likely to admit students from smaller undergraduate schools.

To characterize an applicant's undergraduate background, we used two measures. First, we used Barron's value, which is a measure of an institution's selectivity based on incoming students' SAT scores, GPA and class rank, and overall acceptance rates. While not equivalent to prestigious, we treat selectivity as a proxy for prestige based on the assumption that more selective institutions are also prestigious institutions. For our analysis, we defined institutions with Barron's values of "most competitive" or "highly competitive" as selective and all other institutions as not selective.

Second, we used the number of bachelor's degrees awarded by the physics department at the applicant's undergraduate institution to estimate the size and notoriety of the department, with the assumption that a department that grants more degrees is more likely to be known by an admissions committee member. Due to variability in yearly degrees, we used the median number of degrees over the 2016-2017, 2017-2018, and 2018-2019 academic years as the number of bachelor's degree awarded \cite{nicholson_roster_2018,nicholson_roster_2019, nicholson_roster_2020}. We then defined any program that was in the top quartile of physics bachelor's degrees awarded during that period as a large program and all other programs as smaller programs. For reference, the programs we classified as large produced nearly two-thirds of all physics bachelors degrees over the period.

To perform the comparisons in all cases, we used Fisher's Exact Test to examine whether the rubric score was associated with any of the metrics of interest (admission status, sex, institution selectivity, institution size). We used the standard choice of $\alpha=0.05$ to judge claims of statistical significance. Because we did 18 comparisons for each metric of interest, it is likely that there would be at least one false positive. Therefore, we used the Holm-Bonferroni procedure to correct the p-values for multiple comparisons as it is less conservative than the traditional Bonferroni correction while maintaining statistical power \cite{holm_simple_1979}.

For cases of missing data, we used pairwise deletion so that we could make the most use of the data we had. While Nissen et al recommends using multiple imputations for missing data in physics education research studies \cite{nissen_missing_2019}, the goal of this paper is to understand what faculty did as opposed to estimate a larger trend or predict an outcome. Therefore, we do not believe that using multiple imputations is aligned with the goal of this paper. The percent of missing data per rubric metric is shown in Table \ref{tab:missing_values}.

\begin{table}[]
\caption{Percent of Missing Data by Rubric Construct}
\label{tab:missing_values}
\begin{tabular}{lr}
\hline
Rubric Construct             & Percent Missing \\ \hline
Physics Coursework           & 20.0      \\
Math Coursework              & 20.2      \\
All Other Coursework         & 20.2      \\
Academic Honors              & 22.1      \\
Variety/Duration of Research & 3.4      \\
Quality of work              & 4.4      \\
Technical Skills             & 4.1      \\
Research Dispositions        & 4.7      \\
Achievement Orientation      & 4.4      \\
Conscientiousness            & 4.4      \\
Initiative                   & 4.0      \\
Perseverance                 & 4.4      \\
Alignment of Research        & 7.2      \\
Alignment with Faculty       & 32.1      \\
Community Contributions      & 4.0      \\
Diversity contributions      & 3.4      \\
General GRE Scores           & 2.2      \\
Physics GRE Score            & 2.5      \\ \hline
\end{tabular}
\end{table}

\section{Results}\label{sec:Results}
The results are largely unchanged based on whether we rounded up or rounded down when a faculty member gave a rating in-between levels of the rubric so we present only the rounded up results here.

\begin{figure}
 \includegraphics[width=1\linewidth]{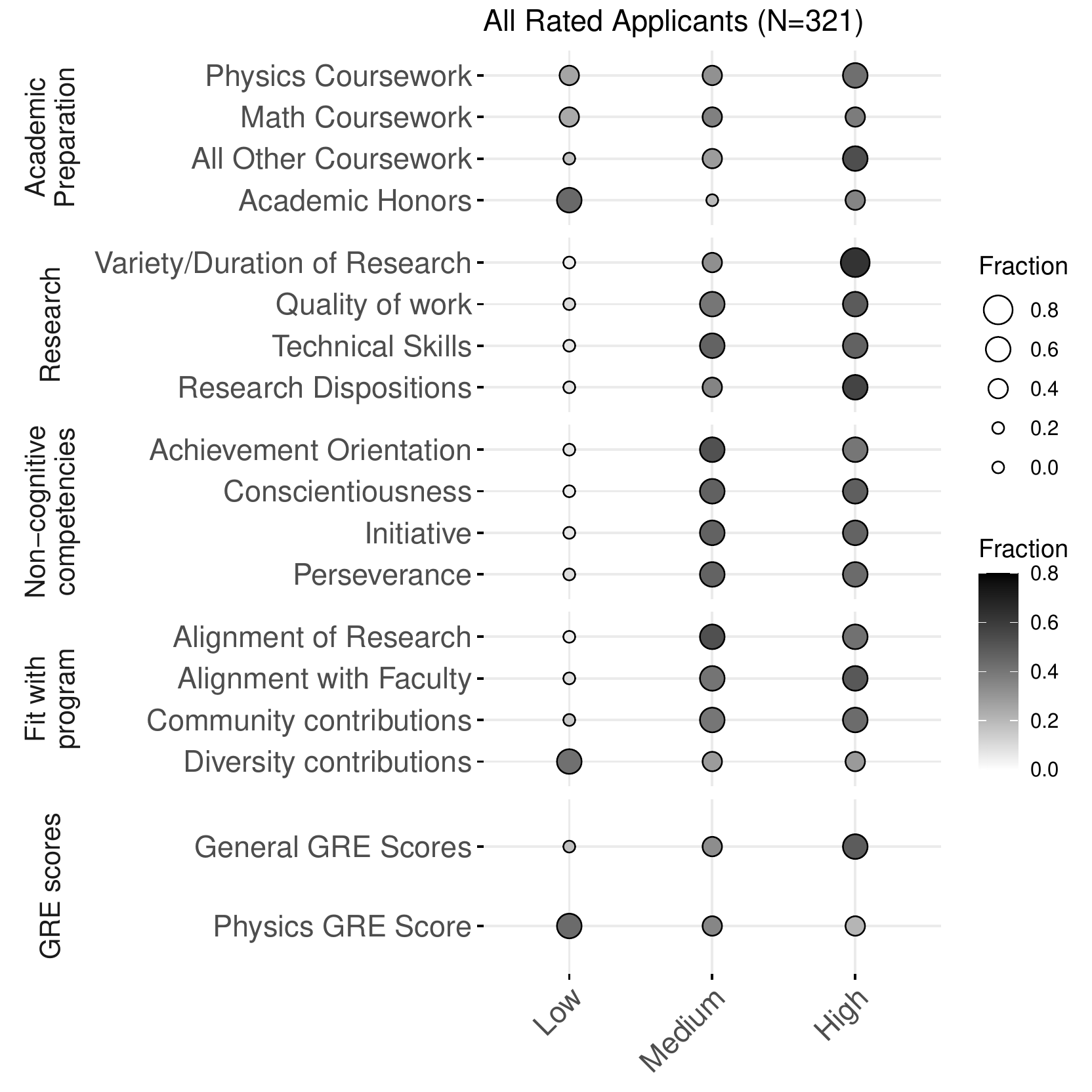}
 \caption{Faculty ratings of domestic applicants on 18 constructs. In the plot, a larger, darker circle means that more applicants are in that bin. While many applicants are in each level of the academic preparation and test score constructs, few applicants are in the "low" bin of the research, noncognitive skills, and program fit constructs.  \label{fig:balloon_all}}
\end{figure}

\begin{figure*}
 \includegraphics[width=1\linewidth]{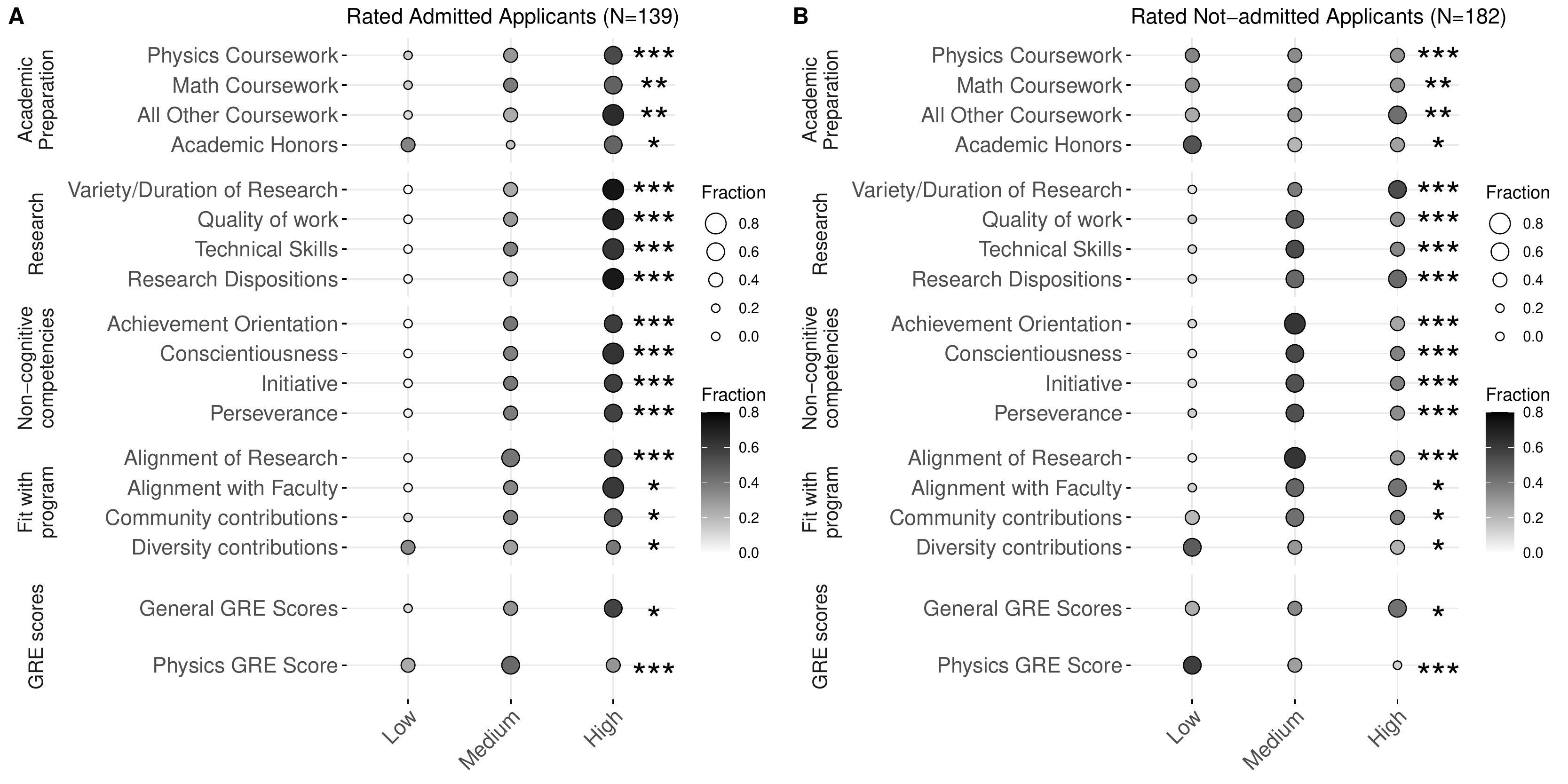}
 \caption{Faculty ratings of domestic applicants on 18 constructs split by whether the applicant was admitted. Ratings that are statistically different between the two plots are marked on the right side of the plot, with a corrected p-value <0.001 represented by "***", <0.01 by "**", and <0.05 by "*". The distribution of ratings of all constructs is statistically different for admitted applicants compared to non-admitted applicants. Overall, most admitted applicants were rated "high" while most non-admitted applicants were rated "medium." \label{fig:balloon_admit}}
\end{figure*}

\begin{figure*}
 \includegraphics[width=1\linewidth]{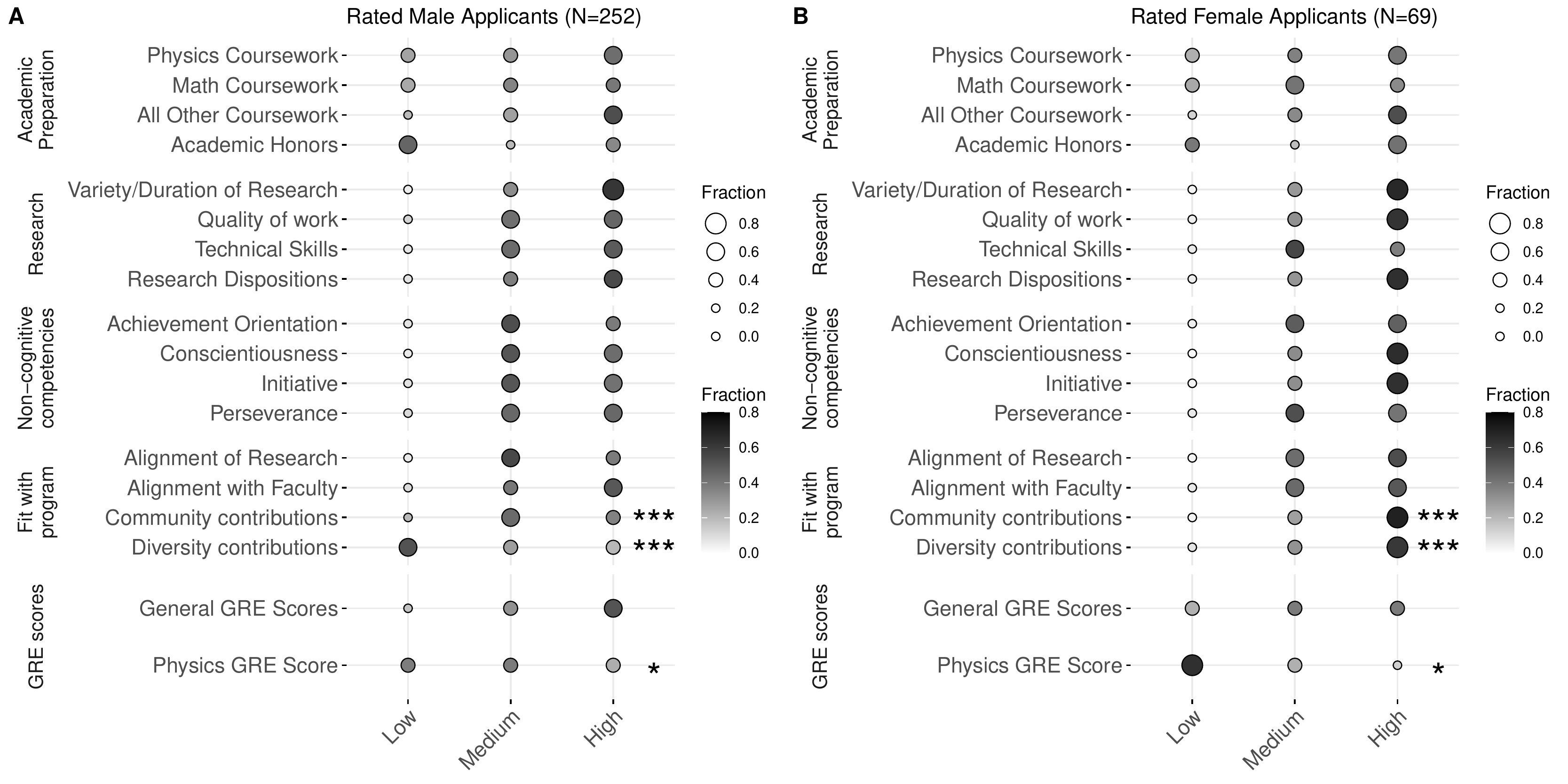}
 \caption{Faculty ratings of domestic applicants on 18 constructs split by whether the applicant was male or female. Ratings that are statistically different between the two plots are marked on the right side of the plot, with a corrected p-value <0.001 represented by "***", <0.01 by "**", and <0.05 by "*".  Only three of the constructs showed statistical differences between males and females: physics GRE score where males scored higher and community contributions and diversity contributions where females scored higher. \label{fig:balloon_gender}}
\end{figure*}

\begin{figure*}
 \includegraphics[width=1\linewidth]{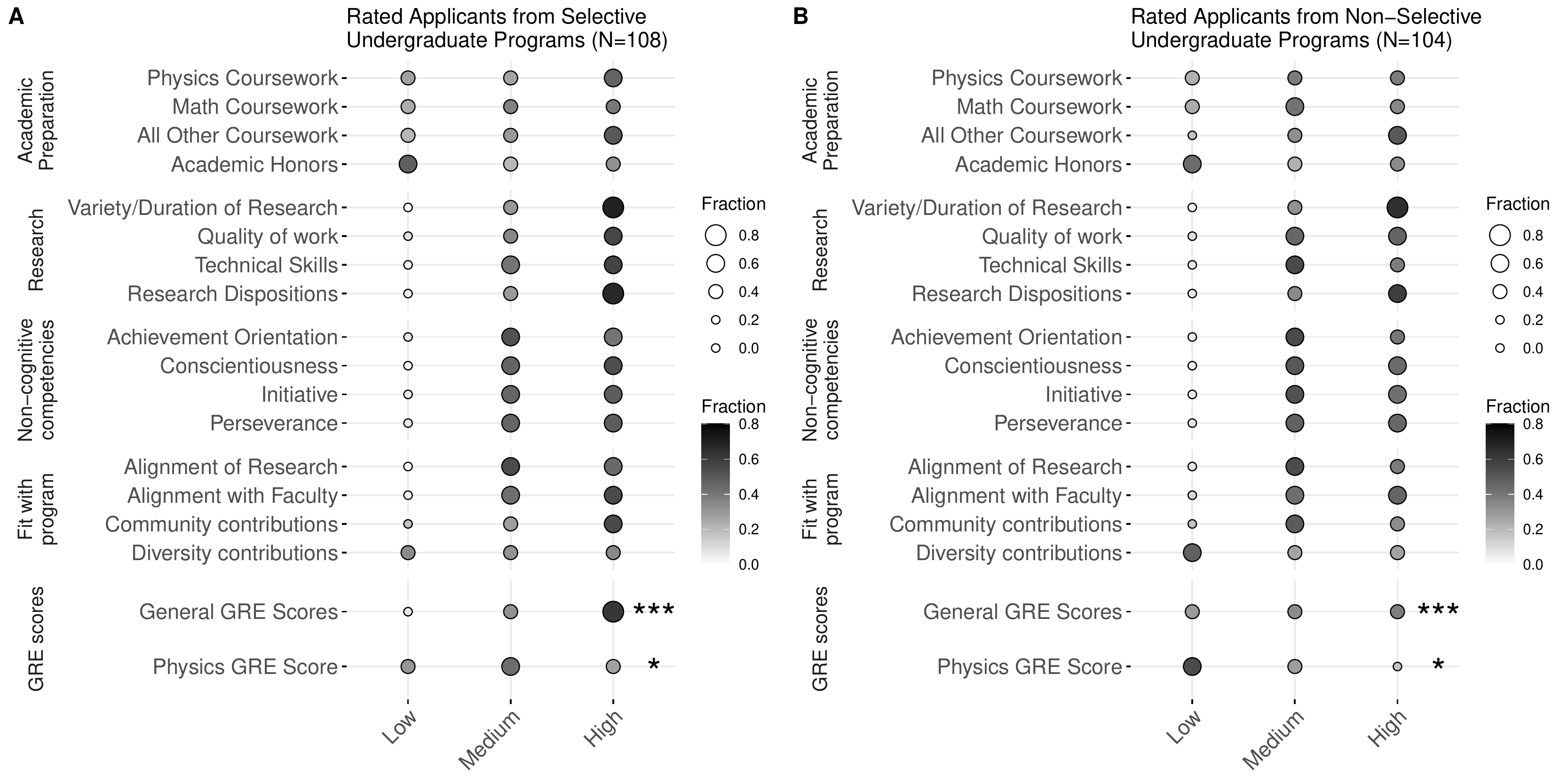}
 \caption{Faculty ratings of domestic applicants on 18 constructs split by whether the applicant attended a more selective or less selective undergraduate university. Ratings that are statistically different between the two plots are marked on the right side of the plot, with a corrected p-value <0.001 represented by "***", <0.01 by "**", and <0.05 by "*". Only the general GRE and physics GRE scores showed differences. \label{fig:balloon_barron}}
\end{figure*}

\begin{figure*}
 \includegraphics[width=1\linewidth]{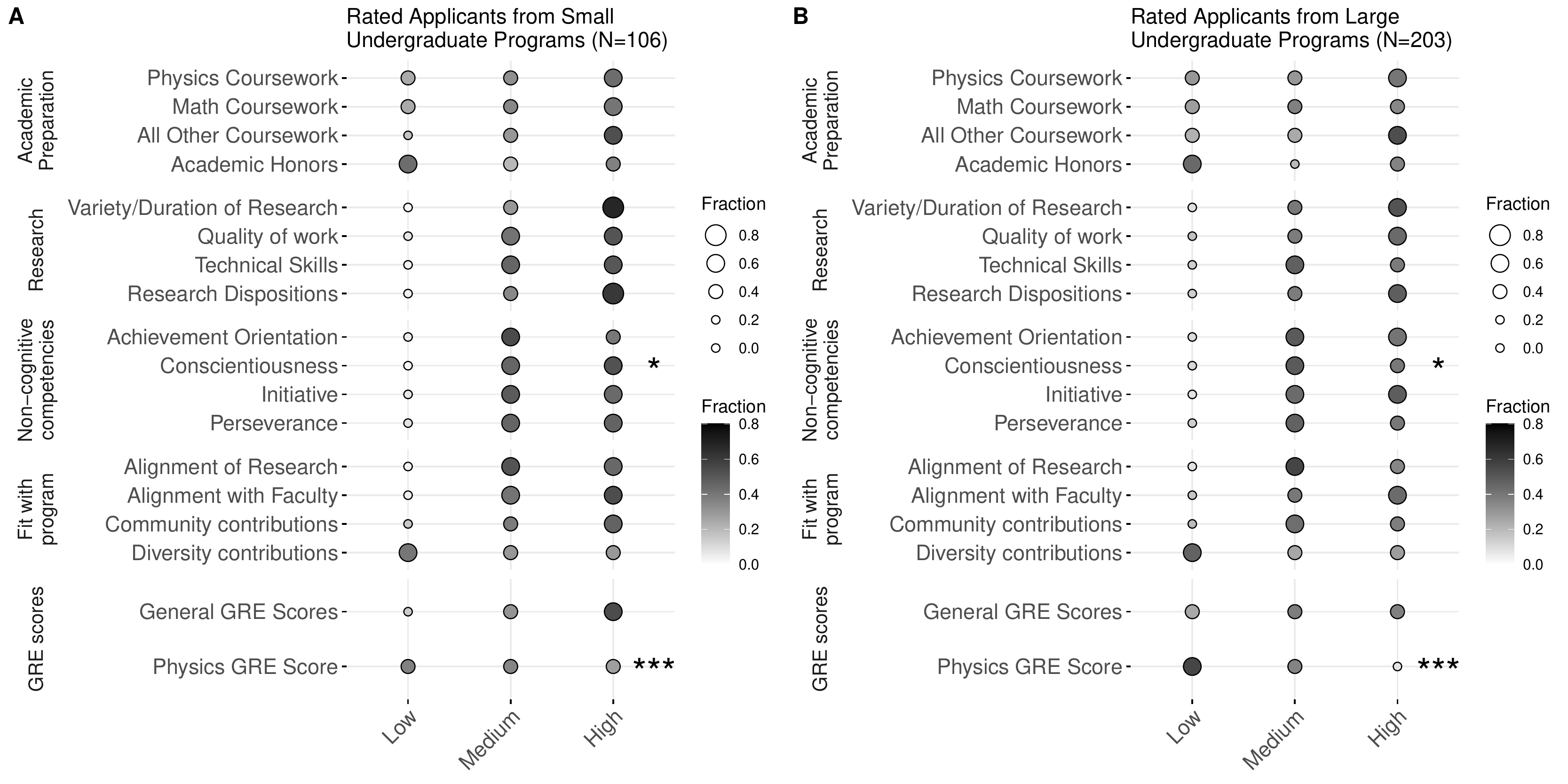}
 \caption{Faculty ratings of domestic applicants on 18 constructs split by whether the applicant attended a university with a larger or smaller physics program. Ratings that are statistically different between the two plots are marked on the right side of the plot, with a corrected p-value <0.001 represented by "***", <0.01 by "**", and <0.05 by "*". Only the physics GRE score and conscientiousness showed differences between the groups of applicants, with the latter dependent on how larger physics program is defined. \label{fig:balloon_size}}
\end{figure*}

When we examine the faculty's rating of all applicants in Fig. \ref{fig:balloon_all}, we notice two overarching trends. First, for traditional measures of academic success such as grades and test scores, faculty tend to rate applicants using all three levels of the rubric. For the academic preparation constructs on the rubric, "high" is the most common rating given by faculty. However in terms of math and physics course grades, around 25\% of applicants still scored in the low bin. Of the academic preparation constructs, academic honors follows a different structure than the others where faculty ratings are bi-modal, meaning that applicants either had no academic honors or had multiple academic honors.

Second, for the research, noncognitive, and fit constructs, faculty rarely used the "low" level of the rubric, with only three of the twelve constructs in those categories having more than 10\% of applicants earning a "low." For research, the most common rating was "high" while for the noncognitive traits, the most common rating varied between "high" and "medium." In terms of the fit constructs, most applicants were rated as either "medium" or "high" for alignment of research, alignment with faculty, and community contributions. In contrast, for the diversity contributions construct, "low" was the most common rating, meaning that many applicants did not discuss how they promote or advocate for diversity in their applications.

When looking at how faculty rate applicants who would later be admitted compared to applicants who would not be admitted, we see statistically significant differences in the distribution of all ratings (Fig. \ref{fig:balloon_admit}). Overall, admitted applicants tended to be rated "high" on each construct while non-admitted applicants tended to be rated "medium" on each construct. There were a few exceptions to the general trend however. For academic honors, diversity contributions, and physics GRE scores, most admitted students were not rated as "High" and 25\% of applicants received a "low" score while for all other course work, variety/duration of research, and general GRE scores, most non-admitted applicants were rated as high.

When looking at the ratings broken down by sex (Fig. \ref{fig:balloon_gender}), we notice that the results tend to follow the overall patterns of all three ratings on academic success and test scores and mainly "medium" and "high" ratings on research, noncognitive skills, and fit with the program for both males and females. Comparing ratings between males and females, we find that only physics GRE score, community contributions, and diversity contributions showed statistically significant differences. While males tended to score higher on the physics GRE score, females tended to score higher on community contributions and diversity contributions. As we elaborate on in the discussion, differences in these three constructs do not necessarily mean that faculty are rating males and females differently but instead may be documenting inequities that already exist.

Likewise, when looking at the ratings broken down by the selectivity of the university where the applicant earned their bachelor's degree (Fig. \ref{fig:balloon_barron} ) or the size of the department where they earned their bachelor's degree (Fig. \ref{fig:balloon_size}), we may also be observing existing inequities reflected in the faculty ratings. For example, applicants from more selective universities only had statistically higher ratings than applicants from less selective universities on the general GRE and physics GRE scores. Similarly applicants from larger programs had statistically higher ratings on the physics GRE score than applicants from smaller programs did. However, applicants from larger programs were also rated higher on conscientiousness than applicants from smaller programs, though this result is sensitive to how we define a large program.

While we could consider interactions between admission status and sex, institutional selectivity, or physics program size, we did not do so given the small sample sizes. For completeness, however, we include those plots in the supplemental material.

\section{Discussion}\label{sec:discussion}
\emph{How do faculty assign rubric scores to applicants and how do those differ between admitted and rejected applicants?}
For academic achievement and test scores, faculty tended to use all three levels of the rubric when assigning scores to applicants. In contrast, faculty tended to use mainly "medium" and "high" when assigning scores to applicants in the research, noncognitive skills, and program fit categories. We argue that this result is more of a reflection of the rubric than it is of how faculty are using the rubric.

Given that grades and test scores are well defined via transcripts and test scores, rubric constructs measuring these tended to use quantitative measures to determine which score the applicant would receive. That is, a high score or high grades would correspond to a "high" rating while a low score or low grades would correspond to a "low" rating. Additionally, as the courses required for a physics degree tend to be similar regardless of the specific program, most applicants will have taken the courses mentioned in the rubric and hence, faculty can rank applicants based on those grades.

In contrast, the research, noncognitive skills, and fit with department are less well defined and instead depend on what applicants write in their statements and what information letters of recommendation contain. This means that not all constructs on the rubric may necessarily be addressed. For example, if an applicant takes quantum mechanics it will certainly appear on their transcript but if that applicant was also active in departmental service activities, it may not be reflected in any parts of the application. As a result, the rubric needs to take into account that applicants may not display a trait because either they do not exhibit it or because they did not mention it (the instructions for the applicant's statements ask them to address multiple topics that map onto various rubric constructs). Any display of the trait could then not fall into the "low" level of the rubric, which would then explain why faculty tended to use only the "medium" and "high" ratings.

A reasonable follow up is then whether combining "no evidence" with "evidence not presented" as a single level on the rubric represents an issue with the rubric. We argue that it does not, as it provides the best option given the data faculty have available. Applicants are asked to discuss certain topics in their statements that map broadly onto the rubric constructs but that does not necessarily mean they will. While interviews could be useful in separating "no evidence" cases from "evidence not presented" cases, we worry these would increase admissions committee members' work load.

In terms of comparing admitted and non-admitted applicants, all 18 rubric constructs showed statistically significant differences. Given the goal of the rubric is to aid faculty in determining who to admit, we would expect the rubric to show such differences. That all rubric constructs show differences suggests all parts of the rubric are useful for determining who to admit.

\emph{How do the scores assigned by faculty differ by applicant’s sex?}
We found only three constructs on the rubric that showed sex differences: physics GRE score, community contributions, and diversity contributions. Given known scoring gaps on the physics GRE \cite{miller_typical_2019}, it is not surprising that males are rated more highly than females on the physics GRE score. Given that females perform larger amounts of service work in academia \cite{guarino_faculty_2017}, it is also not unexpected that constructs measuring these would show a difference between sexes. Because the constructs that show sex differences are related to effects documented in the literature, we believe that the rubric is reflecting inequities that already exist rather than creating additional ones. Therefore, we conclude that the rubric is not providing an advantage to male or female applicants.

Additionally, the constructs of the rubric that do not show differences between sexes also align with what we would expect based on the literature. The result that physics and math GPA did not differ by sex aligns with the findings of \cite{whitcomb_not_2020} and the result that noncognitive skills did not differ by sex aligns with the general finding that noncognitive skills do not appear to depend on demographics \cite{kyllonen_noncognitive_2005,miller_using_2015}.

\emph{How do the scores assigned by faculty differ by the type of institution the applicant attended?}
When we compared applicants based on whether their undergraduate institution was a more or less selective institution, we found that the only constructs that showed differences were the general GRE and physics GRE scores. This result aligns with the results of our previous work investigating the physics GRE scores by undergraduate institution type \cite{young_physics_2020,mikkelsen_investigating_2021}. We note that if we instead define more-selective universities to include large state universities, such as Michigan State University, University of Colorado, Boulder, and University of Washington, our results are unchanged. This redefinition is equivalent to considering Barron's values of 1-3 as more-selective and everything else as less-selective compared to the definition of more-selective as Barron's values of 1 and 2 in the Methods and Results sections.

The interpretation of the results when comparing applicants from larger or smaller physics departments is less straightforward because the results do depend on how we define "larger" and "smaller" departments. When we define larger programs as those that ranked in the top quartile of physics bachelor's degrees granted as measured by the median number of degrees awarded over the last three years of available data and rounded up in-between ratings, we find that the physics GRE score and conscientiousness showed differences between applicants from larger and smaller programs. However, if we rounded down on in-between ratings instead, only physics GRE score showed a difference between applicants from larger and smaller programs.

Furthermore, alternative definitions of "larger programs" also produced varying results. One could also have reasonably defined "larger" to mean 1) in the top half of physics bachelor's degrees granted as measured by the median number of degrees awarded over the last three years, 2) in the top quartile of physics bachelor's degrees granted as measured by the total number of degrees awarded over the last three years, and 3) in the top half of physics bachelor's degrees granted as measured by the total number of degrees awarded over the last three years. When we also consider rounding up or rounding down in-between ratings, we could make various combinations of physics GRE score, general GRE score, physics coursework, and conscientiousness show a statistically significant difference. The only rubric construct that always showed a statistically significant difference regardless of how we defined "larger programs" was the physics GRE score. Therefore, the results suggest that applicants from larger physics programs score higher on the physics GRE than applicants from smaller program do, but the results are inconclusive as to whether other areas of the rubric might show differences based on the size of the physics program the applicant attended.

One area that unexpectedly did not show differences regardless of how we defined "larger program" was the research section. It is often assumed that students at larger programs have more opportunities to engage in research than students at smaller programs. Yet, even if that is true, it does not appear to be reflected in the rubric scores.

\section{Limitations}\label{sec:Limitations}
Our study has four main limitations. First, our study does not include many disadvantaged groups in higher education who might not have the same opportunities as their more privileged peers and hence, may score lower on the rubric. While gender and race are the most obvious due to the way our university records applicant data and interprets Proposal 2, our study does not include a comparison of low-income applicants to higher-income applicants or first generation applicants to continuing generation applicants.

Additionally, the size of our study does not allow us to explore intersections and where possible inequities may lie. As Rudolph et al. noted, using small sample sizes with sub-groups has insufficient statistical power and could lead to invalid inferences \cite{rudolph_final_2020}. Hence, we refrained from performing such analyses in this paper.

Second, our data only contained ratings from the initial reviewer and none of the ratings of later reviews. As a result, we were unable to look into differences in how individual faculty members use the rubric. For example, it is possible that one faculty member might systematically rank applicants lower on the rubric constructs than a different faculty member might. With only a single rating per application, we are unable to disentangle differences in faculty ratings and differences in the applicants the faculty members reviewed.

Third, this study included only a single program. Under a more traditional graduate admissions system, physics has been called a "high consensus" discipline \cite{posselt_inside_2016}, meaning that physics faculty tend to agree on what a "quality" applicant is and therefore, a single department's admissions process would be more or less representative of graduate admissions processes in physics. When switching to rubric-based admissions, we cannot necessarily make that same claim. As our rubric was created based on what faculty value, it is not unreasonable to assume that the results would generalize to other departments that also use rubric-based admissions. However, until such processes are evaluated at other departments, we cannot make such a claim.

Fourth, as a result of using only one program, the applicants are likely not representative of the larger population. The data in this study comes from 1) people who applied to our program and 2) applicants who had a nearly complete application. Thus, if we consider those with an interest in attending physics graduate school as our population, we first selected on those who applied to graduate school, then selected on those who applied to our program, and finally selected on those who provided enough information in their applications for faculty to evaluate. At each step, we are excluding some of the larger population and thus our claims cannot necessarily be expected to hold for the larger population of potential applicants. For example, anecdotal evidence suggests minoritized applicants are more likely to not complete their applications than majoritized applicants are.

\section{Future Work}\label{sec:future_work}
As noted in the limitations, our study compared rubric scores of males and females and applicants from larger or more selective programs with applicants from smaller or less selective programs. Future work could then explore how rubric-based admissions may impact other historically and currently underrepresented groups in physics such as Black, Latinx, or Indigenous applicants. Racism, and specifically anti-Black racism, is still prevalent in physics \cite{brown_iii_commentary_2020,rosa_educational_2016,barber_systemic_2020,dickens_being_2020,prescod-weinstein_making_2020} and therefore might be reflected in rubric-based admissions. 

While physics faculty tend to think of diversity mainly in terms of race \cite{posselt_inside_2016}, we acknowledge that diversity is broader than race and studies of equity around the rubric should also consider first generation applicants, low-income applicants, disabled applicants, and veterans. Studies of undergraduate admissions suggest that when extracurriculars and subjective assessments of character and talent gleaned from essays and recommendations are added to the admissions process, existing inequalities may increase \cite{rosinger_role_2020} and these applicants may become further disadvantaged in the admissions process. Therefore, future work should ensure that rubric-based admissions do increase equity rather than just use a new tool to perpetuate existing inequities.

Second, future work should examine how the use of rubrics may affect what parts of an application drive the admissions process. In our prior work, we found that the physics GRE and grade point average were the main drivers of the admissions process \cite{young_identifying_2019}. Given the rubric is designed to emphasize more than just grades and test scores, we would hope to see these factors deemphasized under the rubric system. Such a result would suggest that the rubric is fundamentally changing how faculty are reviewing applicants.

Third, future work could examine how rubric-based admissions may change the type of applicant admitted and student outcomes. Faculty skeptical of holistic admissions may worry that by deemphasizing grades and test scores, their program is admitting less academically prepared students. Future work can explore if these fears have any merit. Research at the undergraduate level on holistic admissions has found that adding noncognitive traits increased graduation rates, especially among those from disadvantaged backgrounds \cite{kalsbeek_employing_2013}. At the graduate level, a study of a materials science and engineering program found that after changing their admissions to include noncognitive skills, their incoming students won more university fellowships, though the authors cautioned they could not attribute the increase in fellowships solely to their changes in admissions \cite{stiner-jones_2019_2020}. Thus, evidence from outside of physics suggests that these fears may be unfounded, but we will not know for sure until physics specific studies are conducted.

Additionally, future work can examine noncognitive skills in physics more broadly. Physics has been characterized as a brilliance-dominated field \cite{leslie_expectations_2015} and hence, it is not surprising that most studies of success in physics have also focused on cognitive measures such as grades, exam scores, and standardized test scores. While such studies could be useful at all levels of physics, studies at the graduate level are especially important given the limited number of studies exploring their usefulness for predicting success in graduate school. \cite{rudolph_final_2020}.

Finally, future work around equity in graduate admissions should investigate who is invited to apply to graduate school in the first place, what barriers those who do not apply but wish to do so encounter, and how those barriers may be removed. In previous work, Cochran et al. investigated what barriers applicants to physics graduate school, via the APS Bridge Program, perceived, finding that GRE scores, lack of research experience, low GPA, program deadlines, and application costs were common concerns \cite{cochran_identifying_2018}. Unless we also work to make the application process more equitable, making the evaluation process more equitable will not result in large-scale changes in equity at the graduate level.

Shifting from a researcher lens to a practitioner lens, future work can also examine how graduate programs as a whole can become more equitable and how graduate programs can ensure their program goals align with their admissions goals. Milestones in a typical physics graduate program include passing a series of courses and exams followed by completing independent research for a dissertation. While common, such assessment practices should be evaluated as to whether they are aligned with the goals for admission. For example, after implementing rubric-based admissions, our department no longer requires incoming students to take and pass a qualifying exam.

\section{Recommendations for Departments}\label{recommendations}
The results of this study suggest a general recommendation to implement rubrics in physics graduate school admissions. Rubrics can aid reviewing applications by standardizing the process and limiting bias and using rubrics does not appear to increase the time to review applications.

Of course, simply using a rubric will not result in changes unless it is implemented well. We therefore propose three more specific recommendations.

First, we recommend that admissions committees have multiple members review each application. For a well-constructed rubric, there should be limited uncertainty as to what rating an applicant will receive. However, for constructs that are more subjective in nature, faculty may have differing opinions about what counts as achieving each level. For example, for the quality of work construct on our rubric, what counts as "making significant contributions to the project" might vary based on the reviewer. Therefore, having multiple reviewers can reduce potential bias when reviewing applications.

Second, following the call of others \cite{ets_curated_nodate, capers_strategies_2018,roberts_review_2021}, we recommend that members of the admissions committee should be of diverse backgrounds and representative of the applicant pool. To accomplish that, departments might also consider adding non-tenure stream faculty, post-docs, and current graduate students to their admissions committees, providing appropriate recognition and compensation as necessary. Prior work has shown that faculty may prefer to admit applicants like themselves \cite{posselt_inside_2016} and therefore, a representative admissions committee is needed to ensure that minoritized applicants are given equal consideration.

Finally, we recommend that departments conduct regular self-studies of their graduate admissions processes and share the results. While Rudolph et al. have previously called for departments to conduct self-studies of their admissions process \cite{rudolph_final_2020}, we believe it is equally important to share the results of those self-studies so that the physics community can know what is and what is not working. This collective knowledge of what is working and what is not working can then be used by all to improve graduate admissions in physics for everyone.

For the sharing of results to be impactful however, the results must be easy to access and easy to understand. While individual departments could post their results on their websites, we believe doing so adds an extra layer of complexity and makes the results harder to access. Instead, we advocate for a centralized system to be created so that departments can easily report their data in a standardized way and practitioners can easily see and compare results across programs. Such a system could be maintained by professional societies such the American Physical Society or the American Institute of Physics, or other organizations. A system like this has been designed for research-based assessments \cite{nissen_participation_2018}, but to our knowledge, there exists no such system for graduate admissions.

However, when conducting such self-study of what is working well and what is not working well, it is important to consider the question of "working well for whom?". As Razack et al. note, "working well" depends on one's social positioning \cite{razack_beyond_2020} and therefore, a change that works well for applicants of one background may not be working for applicants of a different background. By considering the "for whom?", the physics community can ensure that changes made are for the benefit of all rather than as new methods to continue the existing exclusionary practices in graduate admissions.

\section{Conclusion}\label{conclusion}
In this paper, we demonstrated that rubric-based admissions are a promising avenue for increasing equity in graduate admissions. We showed that faculty ratings of applicant's grades, research experiences, and noncognitive abilities do not differ based on the applicant's sex or undergraduate background. The differences we did observe in faculty ratings could be explained as observing known systematic issues in physics regarding test scores and service work expectations.

Based on the results of this study, we recommend that departments use rubric-based holistic review for their graduate admissions process. Multiple people should review each application and those people should be representative of the applicant pool to limit any bias in the review process. Finally, departments should engage in self-study to see how their graduate admissions process is working and share those results so that the physics community can collectively learn what is working and what is not working in making graduate admissions more equitable.

\acknowledgments{We would like to thank Nicole Verboncoeur and Tabitha Hudson for compiling the data in this project. This project was supported by the Michigan State University College of Natural Sciences and the Lappan-Phillips Foundation.}

\bibliography{references.bib} 

\end{document}


\title{Supplemental Material to Rubric-based holistic review: a promising route to equitable graduate admissions in physics}
\maketitle
Here we present figures showing the results split by admission status and gender, undergraduate institution selectivity, and undergraduate physics program size. Given the relatively small sample sizes, we did not conduct tests of statistical significance. We also include the rubric used by our department.

\begin{figure*}
    \centering
    \includegraphics[width=0.95\linewidth]{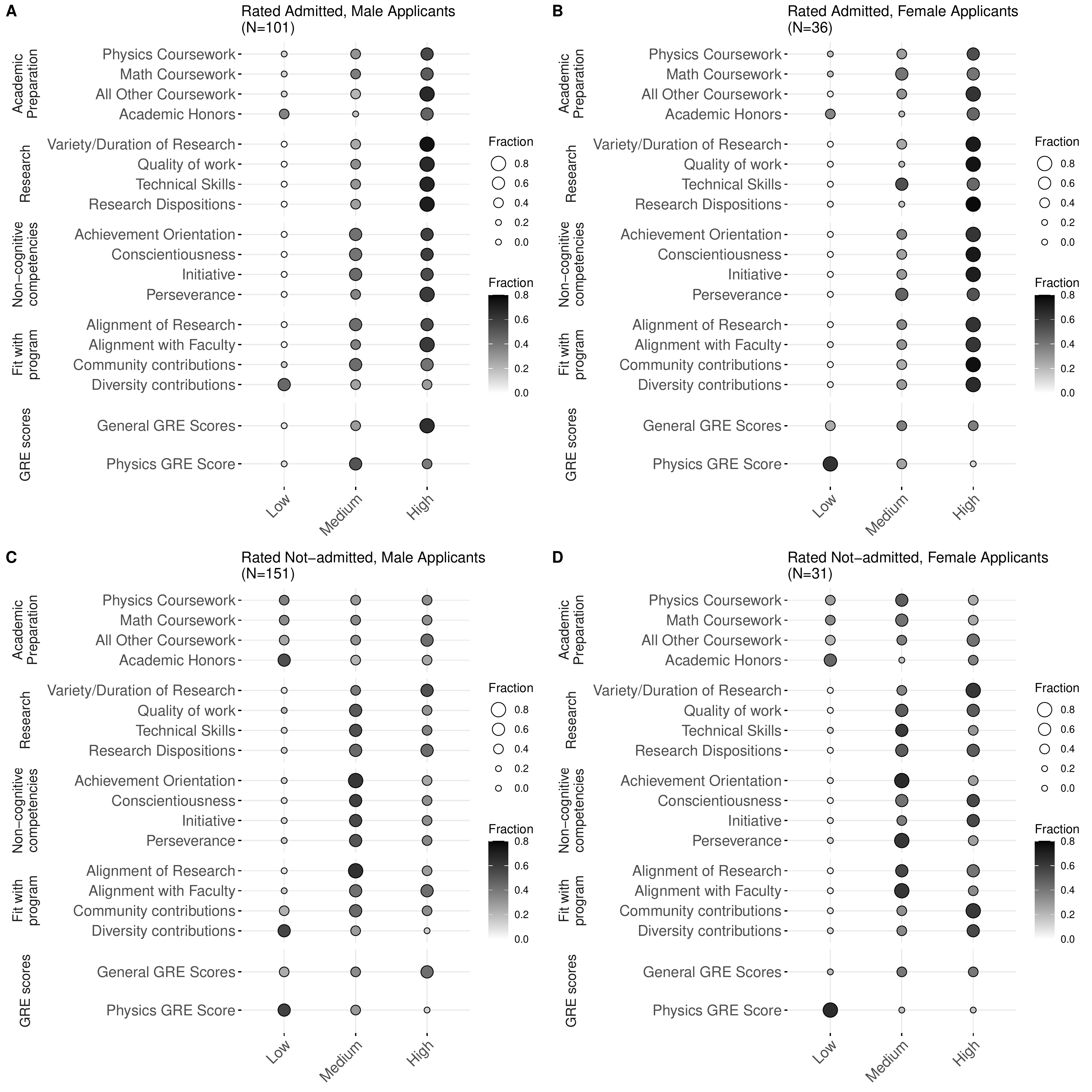}
    \caption{Faculty ratings of domestic applicants on 18 constructs split by whether the applicant was male or female and whether they were admitted or not.}
\end{figure*}

\begin{figure*}
    \centering
    \includegraphics[width=0.95\linewidth]{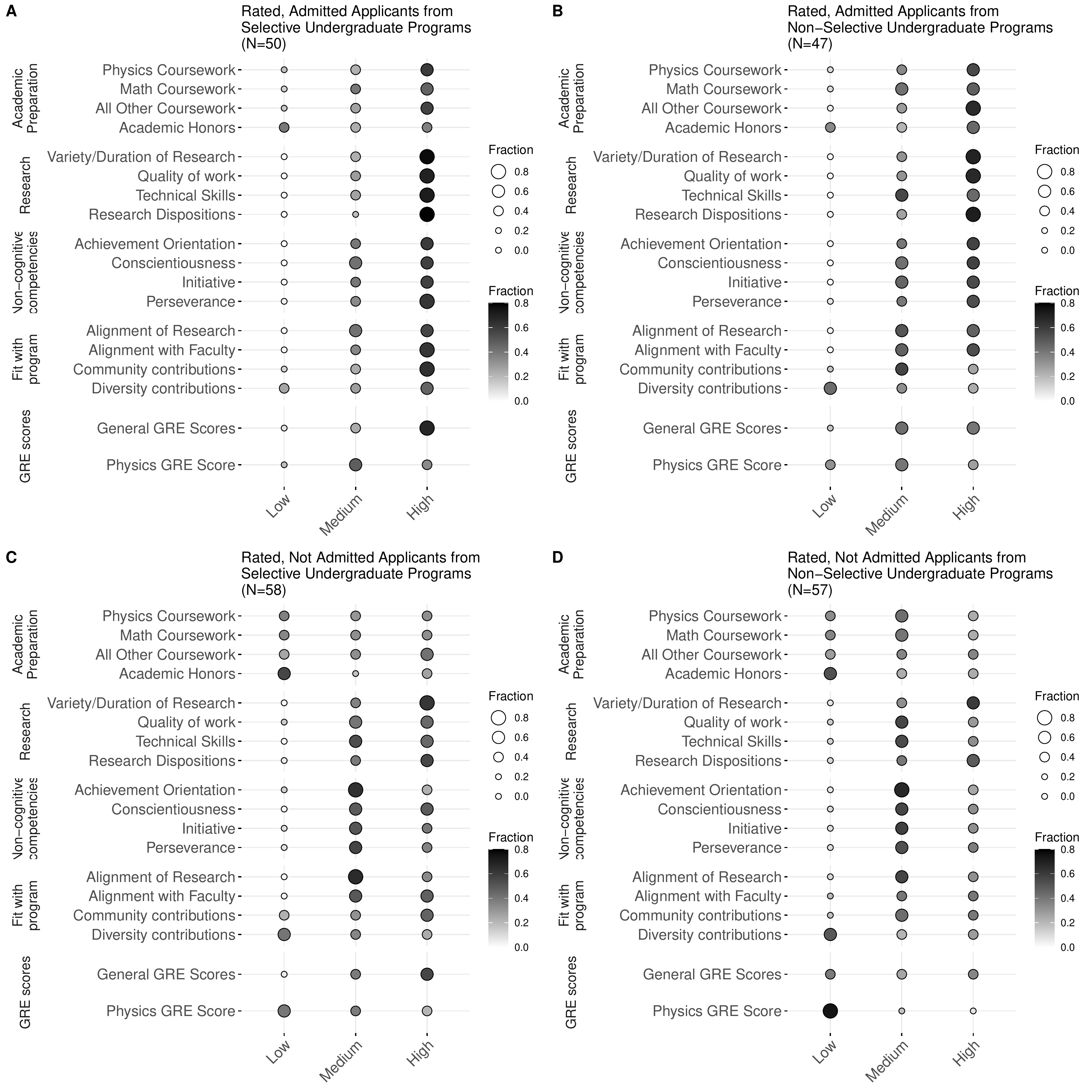}
    \caption{Faculty ratings of domestic applicants on 18 constructs split by whether the applicant attended a more selective or less selective undergraduate university and whether they were admitted or not.}
\end{figure*}

\begin{figure*}
    \centering
    \includegraphics[width=0.95\linewidth]{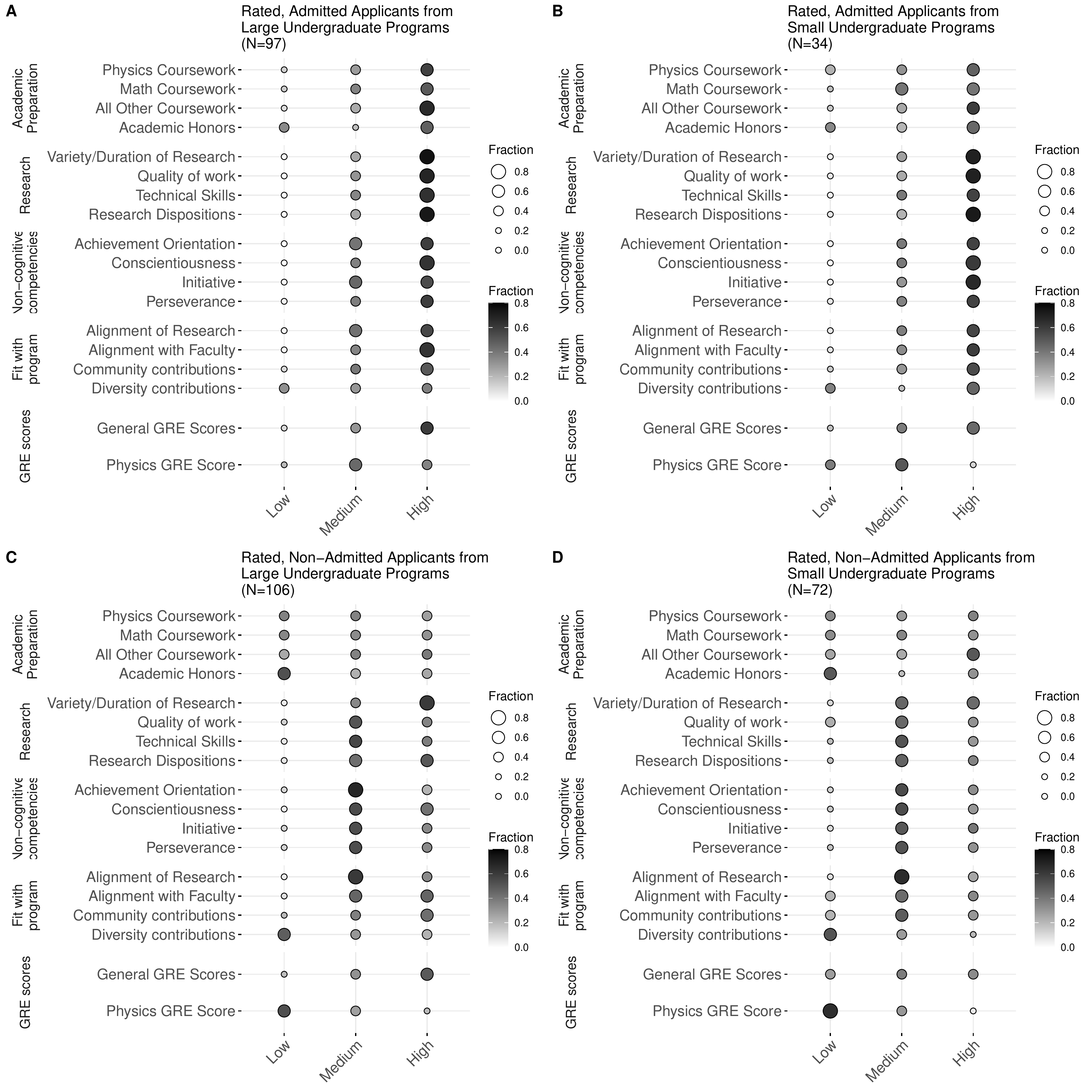}
    \caption{Faculty ratings of domestic applicants on 18 constructs split by whether the applicant attended a university with a larger or smaller physics program and whether they were admitted or not.}
\end{figure*}

\clearpage
\renewcommand*{\arraystretch}{2}
\begin{longtable}{>{\raggedright\arraybackslash}p{1in}>{\raggedright\arraybackslash}p{1in}>{\raggedright\arraybackslash}p{1.5in}>{\raggedright\arraybackslash}p{1.5in}>{\raggedright\arraybackslash}p{1.5in}}
\caption{Rubric used by our department for evaluating applicants. The criteria for a high, medium, and low score are shown.}\label{tab:msu_rubric}\\
\endfirsthead
\caption*{Table \ref{tab:msu_rubric} (cont'd)}\\
\endhead

Item &
  Subitem &
  High &
  Medium &
  Low \\ \hline
Academic Preparation (25\%) &
  Physics Coursework &
  GPA\textgreater{}=3.7 (A-) in all core subjects: CM1\&2, EM1\&2, QM1\&2, SM1, if not taken 2nd semester courses yet are they planning on taking them? &
  GPA \textgreater{}=3.3 (B+) in all core: CM1\&2, EM1\&2, QM1\&2, SM1; OR GPA\textgreater{}=3.7 (A-) in CM1, EM1, QM1, SM1 if no 2nd semester courses taken &
  GPA\textgreater{}=3.7 (A-) in EM1 and CM1; GPA\textgreater{}=3.0 (B) average in other advanced courses; any grades \textless{}2.7 (B-) without explanation \\\hline
 &
  Math Coursework &
  Real and Complex Analysis, Group Theory with GPA\textgreater{}=3.5 (A) grades &
  DiffEq, Linear, and a Math Methods course, all with \textgreater{}=3.5 (A) grades; or more than this with GPA\textgreater{}=3.0 (B or A) grades &
  Bare bones math prep (e.g., up to DiffEq), or low grades regularly on math \\ \hline
 &
  Other Coursework &
  Consistently 3.5 (A) grades &
  Consistently 3.0 (B) grades withnothing below a 2.5 (B-/C+) &
  One or more 2.5s (B-/C+) \\ \hline
 &
  Academic honors and/or recognitions &
  multiple honors, e.g., Dept/University Honors; Phi Beta Kappa, etc &
  one academic award/recognition &
  No academic honors in college documented in the application \\ \hline
Research (25\%) &
  Variety/duration &
  two years in research &
  one year in research; only REUs &
  nothing more than coursework  laboratories \\ \hline
 &
  Quality of work &
  multiple indications of excellence &
  clearly made significant contributions to the project &
  limited intellectual or technical contribution to projects; "button pusher" \\ \hline
 &
  Technical skills &
  a variety of experiment, theory, and/or computational skills &
  has developed only one class of skill (exp or theory or comp) &
  nothing more than coursework laboratories \\ \hline
 &
  Dispositions &
  clear commitment to and enthusiasm for research; AND understands what the process  entails &
  clear commitment to and enthusiasm for research; OR understands what the process  entails &
  not clear if they know what they are getting into with a PhD; seems lukewarm about research \\ \hline
Non-Cognitive Competencies (25\%) &
  Achievement Orientation &
  Consistently strives to improve or  meet a high standard of excellence in all areas &
  Has demonstrated a high standard of excellence in selected areas &
  No evidence of striving for excellence provided in application or student record \\ \hline
 &
  Conscientiousness &
  Takes responsibility for personal performance, both the good and the bad; AND demonstrates efficiency and organization &
  Takes responsibility for personal performance, both the good and the bad; OR demonstrates efficiency and organization &
  No evidence of taking responsibility for performance AND minimal evidence of efficient, organized work \\ \hline
 &
  Initiative &
  Consistently seeks out or acts on opportunities AND takes leadership &
  Consistently seeks out or acts on opportunities OR takes leadership &
  Has not sought out or taken advantage of opportunities AND does not have a record of leadership \\ \hline
 &
  Perseverance &
  Application clearly describes successful coping with failures/obstacles &
  Basic or perfunctory description of overcoming challenges &
  Application does not describe experience with failure/obstacles \\ \hline
Fit with program (15\%) &
  Research &
  research interests align with multiple faculty in multiple subfields &
  research interests align with multiple faculty in one subfield &
  limited alignment between student interests and faculty expertise \\ \hline
 &
  Faculty &
  someone wants to hire as RA now and/or there is a clear fit with current faculty expertise &
  someone could supervise, but interests do not directly support a faculty member's work &
  faculty aligned with applicant's interests are not seeking students \\ \hline
 &
  Community &
  has clearly contributed positively to prior department/school culture, and would do the same for our program &
  some evidence of participating in service activities &
  applicant only discusses him/herself; no evidence of engagement in department or university activities \\ \hline
 &
  Diversity &
  applicant has been an active advocate for diversity in physics &
  belongs to an underrepresented identity group; first generation in college or low SES; and/or contributes to another type of diversity the department seeks &
  contributions to diversity are unclear from the application \\ \hline
GRE Scores (10\%) &
  General GRE &
  Verbal(V) and Quantitative(Q) scores \textgreater{}=75\% (or 157 for V and 160 for Q) AND Analytical Writing (AW) \textgreater{}=4.0 &
  V \& Q scores \textgreater{}=75\% (or 157 for V and 160 for Q) BUT AW\textless{}4.0 &
  V or Q score \textless{}75\% and AW\textless{}4.0 \\ \hline
 &
  Physics GRE &
  \textgreater{}=75\% &
  50-74\% &
  \textless{}49\% \\ \hline
\end{longtable}